\documentclass[12pt]{article}

\usepackage{amsmath,amsfonts,amssymb}
\usepackage{graphicx}
\usepackage{psfrag}
\usepackage{enumerate}

\makeatletter \@addtoreset{equation}{section}

\makeatletter\renewcommand\section{\@startsection {section}{1}{\z@}%
                                   {-3.5ex \@plus -1ex \@minus -.2ex}
                                   {2.3ex \@plus.2ex}%
                                   {\normalfont\large\bfseries}}
\renewcommand\subsection{\@startsection{subsection}{2}{\z@}%
                                     {-3.25ex\@plus -1ex \@minus -.2ex}%
                                     {1.5ex \@plus .2ex}%
                                     {\normalfont\bfseries}}

\newcommand{\be}{\begin{equation}}
\newcommand{\ee}{\end{equation}}
\newcommand{\beq}{\begin{eqnarray}}
\newcommand{\eeq}{\end{eqnarray}}

\def\[{\left [}
\def\]{\right ]}
\def\({\left (}
\def\){\right )}

\def\r2{\sqrt{2}}

\def\n{{\bf \hat{n}}}

\newcommand{\bbibitem}[1]{\bibitem{#1}\marginpar{#1}}

\newcommand{\figref}[1]{Fig. \ref{#1}}
\newcommand{\secref}[1]{Sec. \ref{#1}}

\def\Label#1{\label{#1}%
  \smash{\hbox to0pt{\raise1ex\hbox{\tiny[#1]}\hss}}}
\def\noLabels{\let\Label=\label}
\def\nobbibitem{\let\bbibitem=\bibitem}

\begin{document}
\nobbibitem 

\begin{titlepage}


\vfil\

\begin{center}

{\Large \bf Watching Worlds Collide: \\[.2cm] \large Effects on the
CMB from Cosmological Bubble Collisions}

\vspace{3mm}

Spencer Chang$^{a,b,}$\footnote{e-mail:
spchang123@gmail.com}, Matthew Kleban$^{a,}$\footnote{e-mail: mk161@nyu.edu} and Thomas S.
Levi $^{a,}$\footnote{e-mail: tl34@nyu.edu}
\\

\vspace{5mm}

\bigskip\medskip
\centerline{$^a$\it Center for Cosmology and Particle Physics}
\smallskip\centerline{\it Department of Physics, New York University}
\smallskip\centerline{\it 4 Washington Place, New York, NY 10003.}
\bigskip\centerline{$^b$\it Department of Physics, University of California}
\smallskip\centerline{\it 1 Shields Avenue, Davis, CA 95616.}


\vfil

\end{center}
\setcounter{footnote}{0}
\begin{abstract}
\noindent
We extend our previous work on the cosmology of Coleman-de Luccia bubble collisions. Within a set of approximations we calculate the effects on the cosmic microwave background (CMB) as seen from inside a bubble which has undergone such a collision.  We find that the effects are always qualitatively similar---an anisotropy that depends only on the angle to the collision direction---but can produce a cold or hot spot of varying size, as well as power asymmetries along the axis determined by the collision.  With other parameters held fixed the effects weaken as the amount of inflation which took place inside our bubble grows, but generically survive order 10 efolds past what is required to solve the horizon and flatness problems.  In some regions of parameter space the effects can survive arbitrarily long inflation.

\end{abstract}
\vspace{0.5in}

\end{titlepage}
\renewcommand{\baselinestretch}{1.05}  
\tableofcontents

\newpage

{\it ``Through telescopes men of science constantly search the infinitesimal corners of our solar system seeking new discoveries, hoping to better understand the laws of the Universe.''}

\begin{flushright}
-- When Worlds Collide (The Movie)
\end{flushright}

\section{Introduction}

Perhaps the greatest  challenge for string theory is to find a testable prediction which can differentiate it from other potential theories of quantum gravity.  This has proven to be extremely difficult, because the scale at which string theory effects become important is believed to be far out of reach of the energies accessible to any earth-based particle accelerator.  However cosmology provides another opportunity:  the early universe was arbitrarily hot and dense, and remnants of the physics of that time remain in the large-scale structure of the universe today.

It has become increasingly clear in recent years that one of the characteristic features of string theory is its so-called landscape \cite{discretum,kklt,lennylandscape}.  The landscape is a potential energy function of the would-be moduli scalar fields, which possesses many distinct minima.  These minima are meta-stable and can decay via bubble nucleation-type tunneling or quantum thermal processes.  In the early universe, one expects all the minima to be populated as the temperature decreases and the fields begin to settle.  The minima with the highest values of potential energy then begin to inflate, and are expected to rapidly dominate the volume.  After a few string times the universe will be almost completely full of rapidly inflating regions, and as time continues to pass the metastable field values in these regions will begin to decay by bubble nucleation.  Since we live in a region with very small vacuum energy, we expect that our observable universe is contained in such a bubble.

The most promising way to test this idea is to look at large scale cosmological structure.  If we model the string theory landscape as a scalar field coupled to gravity with a potential with several minima, universes inside a bubble which formed via a tunneling instanton have certain characteristic features:  they are open, they are curvature dominated at the ``big bang" (rendering it non-singular), and they have a characteristic spectrum of density perturbations generated during the inflationary period that takes place inside them  \cite{cdl,gott1,gott2,Garriga:1998he,lindeopen,Freivogel:2005vv,batra}.

Moreover, because the bubble forms inside a metastable region, other bubbles will appear in the space around it.  If these bubbles appear close enough to ours, they will collide with it \cite{oldguth,ggv,Aguirre:2007an}.  The two bubbles then ``stick" together, separated by a domain wall, and the force of the collision releases a pulse of energy which propagates through each of the pair.  This process (as seen from inside our bubble) makes the universe anisotropic, affects the cosmic microwave background and the formation of structure.  A collision between two bubbles produces a spacetime which can be solved for exactly \cite{ben,wwc,Aguirre:2007wm}.  

In \cite{wwc}, we began the study of this process.  We were primarily interested in the effects on the cosmic microwave background (CMB).  Our solution shows that the CMB sky will be divided into two (unequal) halves along a circle.  The size of the circle depends on when the collision took place.  The temperature fluctuations in the disk on the side of the circle away from the collision will be weakly affected (by the gravitational backreaction of the energy dumped into our bubble) in a relatively simple way, which we calculated.  The disk on the other side will be more strongly affected, and in a way that depends in a more complex way on the model which we did not calculate.  In this paper we will use a simplified model to compute these latter effects.

Our model consists of a collision between our bubble---which we treat as a thin-wall Coleman-de Luccia bubble which undergoes a period of inflation after it formed (which we model as a de Sitter phase) followed by a period of radiation domination---and another different bubble, the details of which are largely unimportant.  We assume that the tension of the domain wall between the bubbles and the vacuum energy in the other bubble are such that the domain wall accelerates away from us.  Such collisions can produce observable effects which are not in conflict with current data.  We will also assume that the inflaton is the field that underwent the initial tunneling transition, so that the reheating surface is determined by the evolution of that field and can be affected by the collision with the other bubble through its non-linear equation of motion, and that no other scalar fields are relevant.  

After some number of efolds of inflation, our bubble will reheat and become radiation dominated. The subsequent expansion and the geometry of the reheating surface itself will be modified by the presence of the domain wall.  The effect we will focus on is the change in the reheating surface brought about by the domain wall.

There are a number of anomalies in current cosmological data, including a cold spot \cite{coldspot1, coldspot2, coldspot3, coldspot4, coldspot5}, hemispherical power asymmetry \cite{Eriksen:2007pc}, non-Gaussianities (see {\it e.g.} \cite{nongauss}), and the ``axis of evil" \cite{Land:2005ad}.  While the significance of these effects is unclear, there are few if any first-principle models which can account for them.  Our results indicate that bubble collisions naturally create hemispherical power asymmetries spread through the lower CMB multipoles and/or produce relatively small cold or hot spots, depending on the time of the collision (and hence the size of the disk), and can lead to measurable non-Gaussianities.  The qualitative features, however, are always the same:  a collision with a single other bubble produces anisotropies that depend only on one angle---the angle to the vector pointing from our location to the center of the other bubble---and which are generically monotonic in the angle.  
While these results are suggestive, further analysis is required to determine which of these anomalies can be explained by bubble collisions.

While we will not study structure formation in this paper, it is interesting to note that bubble collisions may also create coherent peculiar velocity flows for large-scale structures across large swaths of the sky.  This may be interesting in light of the recently published results \cite{darkflow}.

\section{The Scenario}

In this section we will set up the basic scenario for the bubble collision we seek to study. The details of the dynamics of general collisions were studied in \cite{wwc} and we refer the reader there for a more complete treatment.  We will focus on a collision where our bubble is dominated by positive vacuum energy when it forms, so that it undergoes a period of inflation, and we will assume the pressures are such that the domain wall is accelerating away from the center of our bubble at late times. This second assumption is valid for collisions between two dS bubbles if ours has a smaller cosmological constant.\footnote{It also occurs when our bubble has a larger cosmological constant or in collisions with AdS bubbles when certain conditions involving the tension of the wall are satisfied \cite{ben, wwc, Aguirre:2007wm,Cvetic:1996vr}.}

Because of the difficulty in finding analytic solutions to the full problem of a scalar field with a non-linear potential coupled to gravity, we will make a number of simplifying assumptions.  We enumerate them here:  

\begin{itemize}
\item Unlike in \cite{wwc}, we ignore gravitational backreaction from the shell of radiation released into our bubble by the collision and focus instead on the effects on the inflaton's evolution and reheating.
\item We approximate the inflaton potential as linear when computing its evolution inside our bubble.
\item We allow the field's derivative to be discontinuous across the edge of the lightcone of the collision to model the effects of non-linearities.
\item We treat the wall separating our bubble from the false vacuum, and the domain wall between our bubble and the other bubble, as Dirichlet (fixed value) boundary surfaces for the inflaton.
\item We treat the walls and the edges of the collision lightcone (where the field derivatives change) as thin.
\item We assume inflationary perturbations are unaffected by the collision.
\item We model the effects of the collision on the CMB by computing the redshift to the reheating surface as a function of angle, rather than evolving the fluctuations in the modified spacetime.
\end{itemize}
We will discuss some of these assumptions in more detail later.

\subsection{The Bubble Collision}
In this section we will review the bubble collision spacetime.  Readers familiar with the details of \cite{wwc} may wish to skip ahead to \secref{sec-reheating}.  

A single Coleman-de Luccia bubble in four dimensions has an $SO(3,1)$ invariance (inherited from the $SO(4)$ symmetry of the Euclidean instanton \cite{cdl}). When another bubble collides with ours a preferred direction is picked out, breaking the symmetry down to $SO(2,1)$.  As a result, one can choose coordinates in such a way that the metric takes the form of a warped product between two-dimensional hyperboloids ($H_2$) and a 2D space.

Under the assumption that all the energy in the spacetime remains confined to thin shells or is in the form of vacuum energy, one can find the solution describing the collision by patching together general solutions to Einstein's equations in vacuum plus cosmological constant with this symmetry \cite{ben,wwc, Hawking:1982ga}.  	 The overall result is that the spacetime inside such a bubble is divided into two regions---one that is outside the future lightcone of the collision and can be described using an $SO(3,1)$ invariant metric (a standard open Robertson-Walker cosmology), and one which is inside, is affected by backreaction, and is most conveniently described using 2D hyperboloids.  

\begin{figure}
\centering \hspace{0.2in}
\includegraphics[width=0.6\textwidth]{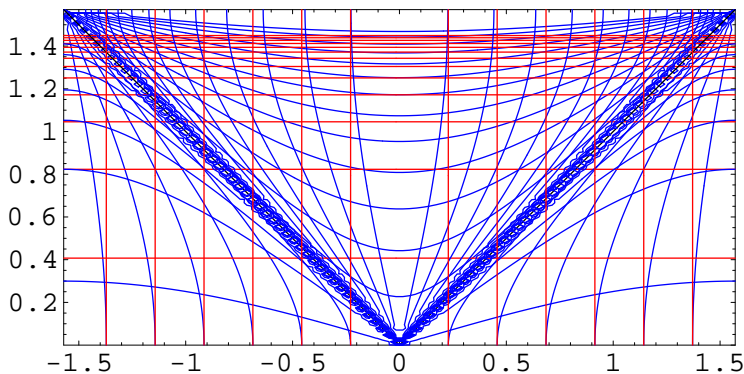} \phantom{xxxxxx}\caption{Plot showing the two coordinates systems for dS space.  The straight (red) lines are $t,x$, and the curvy (blue) lines are $\tau, \xi$.  The values on the axes are in conformal coordinates where the vertical axis, $\tan^{-1}{t}$, runs from 0 to $\pi/2$ and the horizontal axis, $x$, from $-\pi/2$ to $\pi/2$ (with $\ell=1$).
\label{coords}}
\end{figure}

\paragraph{de Sitter space}
The general $SO(2,1)$ invariant spacetime with a positive cosmological constant is given by (see \cite{wwc} for a full discussion of the geometry and causal structure)%
\beq \label{dsback}
ds^2 = {-dt^2 \over g(t)}+g(t)\, dx^2 + t^2 dH_2^2 ,%
\eeq%
where%
\beq%
g(t) = 1+{t^2 \over \ell^2}-{t_0 \over t} ,%
\eeq%
and
\beq%
dH_2 ^2 = d\rho^2 + \sinh^2 \rho d\varphi^2 %
\eeq%
is the metric on the unit 2-hyperboloid and $x \simeq x+ \pi \ell$. We will find for most collisions that $t_0 \ll t$ at cosmologically relevant times and thus we approximate $t_0 \approx 0$, giving $g(t) \approx 1+t^2/\ell^2$. This is a metric for pure dS space with radius of curvature $\ell$. Hereafter, we set $\ell=1$ (it can be easily restored with dimensional analysis). These coordinates (which we denote the $H_2$ coordinates) are related to the $SO(3,1)$  or $H_3$ invariant ones by%
\beq \label{dstrans}%
\cosh \tau &=& \sqrt{1+t^2} \cos x , \nonumber \\
\sinh\tau \sinh\xi \cos \theta &=& \sqrt{1+t^2} \sin x , \nonumber \\[-.3cm]
\\[-.3cm]
\sinh \tau \cosh \xi &=& t \cosh\rho , \nonumber \\
\sinh \tau \sinh \xi \sin \theta &=& t \sinh \rho , \nonumber \\
\varphi &=& \varphi \nonumber ,%
\eeq%
which gives a metric%
\beq%
ds^2 &=& -d\tau^2 + \sinh^2 \tau dH_3^2 , \nonumber \\[-.3cm]
\\[-.3cm]
dH_3^2 &=&  d\xi^2 + \sinh^2 \xi (d\theta^2 + \sin^2 \theta d\varphi^2) \nonumber.
\eeq%
We display the two coordinate systems for dS space in \figref{coords}.

\paragraph{Radiation domination}
We will also need the metric for a radiation dominated (RD) universe. The general $SO(3,1)$ invariant Friedmann-Robertson-Walker (FRW) metric is given by%
\beq%
ds^2=-d\tau^2 + a^2(\tau) dH_3^2 .%
\eeq%
The transformation to $H_2$ coordinates is%
\beq \label{rdtrans}%
a(\tau) \sinh \xi \cos \theta &=& z , \nonumber \\
a(\tau) \sinh \xi \sin \theta &=& t \sinh \rho , \nonumber \\[-.3cm]
\\[-.3cm]
a(\tau) \cosh \xi &=& t \cosh \rho , \nonumber \\
\varphi&=& \varphi . \nonumber %
\eeq%
For radiation domination we have $a(\tau) = \sqrt{C \tau}$, with $C$ a constant. This gives a metric in $H_2$ coordinates%
\beq \label{rdmetric}%
ds^2 &=& { -4 t^4 + (C^2+4t^2)z^2 \over C^2 (t^2-z^2) } dt^2 + {C^2 t^2 - 4z^2 (t^2-z^2) \over C^2 (t^2-z^2) } dz^2 \nonumber \\[-.3cm]
\\[-.3cm]
&&+ {2tz \over C^2 (t^2-z^2)} [ 4 (t^2-z^2)-C^2 ] dt dz +t^2 dH_2^2 . \nonumber%
\eeq%
The radiation dominated part of the spacetime will be affected by the domain wall and the collision. Because of the reduced symmetry (relative to de Sitter), we are currently unable to write down the most general radiation dominated solution with $SO(2,1)$ invariance.\footnote{The problem is equivalent to finding the smooth metric describing a Schwarzschild black hole embedded in an asymptotically FRW spacetime.} However, we expect that at least in a broad class of scenarios the effects on the CMB due to backreaction will be subleading to the other effects we will study, and so we will use this form for now.  It would be interesting to consider these metric effects in the future, for example by solving the equations numerically.  Such a study is currently in progress \cite{wq}.

\subsection{The Reheating Surface \label{sec-reheating}}
\begin{figure}
\centering \hspace{0.2in}
\includegraphics[width=0.5\textwidth]{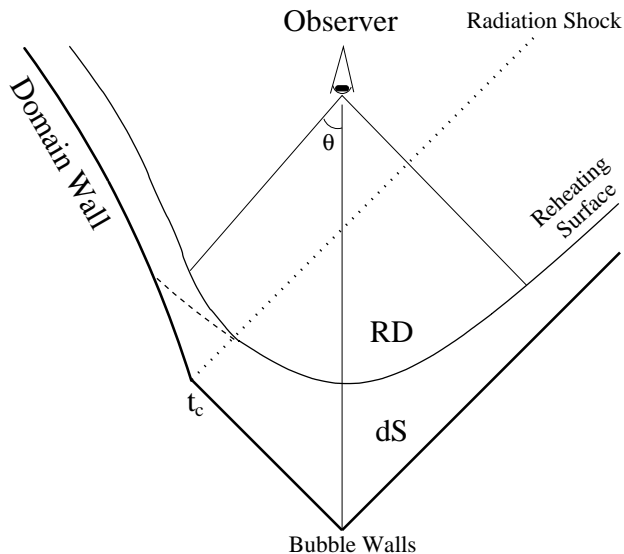}\phantom{xxxxxx} \caption{Sketch of the collision scenario.  Due to the domain wall, the shape of the reheating surface deviates from that of the no collision case (displayed in the long dashed line).   
\label{scen}}
\end{figure}

With these building blocks we can construct our model. Our universe appears as a de Sitter bubble that collides with another bubble, forming a domain wall along the interface.  After the initial period of inflation the universe reheats and transitions to radiation domination. We model this transition as sharp, matching the metrics and Hubble constant across the reheating surface. We ignore matter domination in calculating the effects of the collision on the CMB, and further assume that the effects on the surface of last scattering are related to the effects on the reheating surface simply by the redshift along ``free-streaming'' null geodesics.  In other words, we assume the angular dependence of the redshift to the last scattering surface due to the effects of the collision is given by the angular dependence of the redshift to the modified reheating surface.  
Ignoring the evolution of perturbations between reheating and last scattering is a valid approximation on angular scales larger than that of the first acoustic peak in the CMB (about one degree).  On smaller scales there will be a ``smearing" due to both this evolution and to the finite thickness of the last scattering surface.  Due to these effects, we will focus on disks considerably larger than this size.  We will also ignore the effects of the collision on the perturbations generated during inflation.  While this should be justified when the overall effects are small enough to be consistent with observation, it would be interesting to study this more carefully.  To summarize, in our setup, the main effect is to modulate the standard inflationary perturbations at last scattering by a function dependent on the angle to the collision direction. We sketch the scenario in \figref{scen}.

We work in the thin wall approximation throughout, treating the transition between the two parts of the sky (corresponding to the regions of last scattering which are inside and outside the collision lightcone) as sharp.  It would be interesting to extend the present work to thick wall systems as effects such as the finite thickness of the domain walls and the shell of radiation emitted from the collision point could play interesting roles. 

In a standard Robertson-Walker open universe, reheating occurs on a surface of almost constant density that is $SO(3,1)$ invariant (i.e., hyperbolic 3-space).  During inflation the surfaces of constant energy density are also surfaces of constant value for the inflaton field, and the reheating surface corresponds to the point in the inflaton potential where slow roll comes to an end. This surface defines a preferred reference frame, since the radiation and matter fluids will be at rest on average with respect to it. In our scenario, the collision and domain wall break this symmetry. There is no longer a preferred frame and thus the standard notion of ``comoving'' no longer holds.  There remains a remnant of this symmetry:  a two-dimensional set of comoving observers related to each other by transformations which leave the two-dimensional hyperboloids invariant.  Observers outside this plane are no longer equivalent.

\begin{figure}
\centering \hspace{0.2in}
\includegraphics[width=0.8\textwidth]{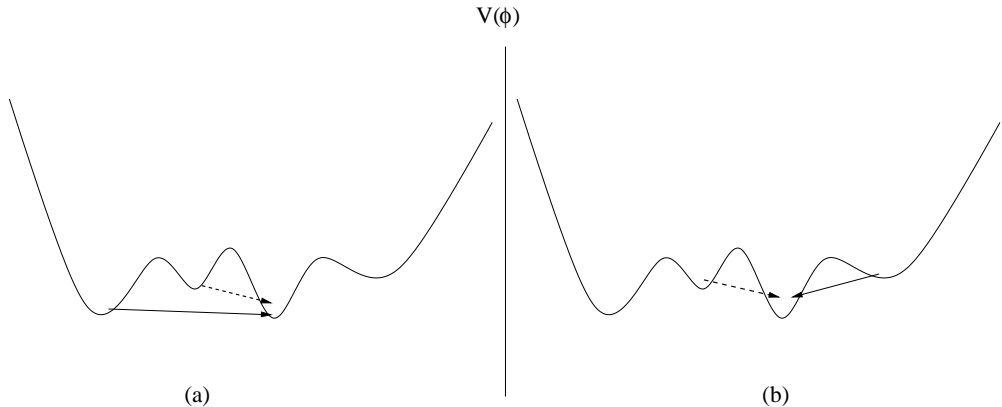} \caption{The scalar field transitions from some vacuum to ours along both the domain and bubble walls. It is possible for the field to transition from the same direction along both (a), or from different directions for each (b).
\label{potential}}
\end{figure}

We define the reheating surface in our scenario as a surface of constant scalar field, $\phi(x,t) = \phi_0$. The presence of the domain wall will alter the shape of this surface in a non-trivial way, breaking isotropy. To fully solve this problem we would need to solve the non-linear coupled scalar field and gravity system for the collision spacetime. Instead we will make an approximation. We expect the scalar field to have a set of values $\phi_L$ in the other bubble, some other set of values $\phi_R$ in our bubble and a third set of values in the metastable vacuum outside both bubbles $\phi_M$. In the thin wall limit the scalar field jumps from near some other local minimum (either the false vacuum our bubble tunneled from, or the vacuum inside the bubble it collided with) to a location a bit up the ``hill'' from our local minimum. This transition occurs all along the domain and bubble walls, so the scalar field is constant along each of these surfaces. 

Without loss of generality we choose $\phi=0$ along the bubble walls. At the domain wall, we have two options. The field could transition at the collision and domain wall from the same direction in field space as the tunneling that initially created our bubble \figref{potential}(a). In this case, to a good approximation we have $\phi=0$ along the domain wall as well. However, it is also possible for the field to transition from the other side of the potential \figref{potential}(b), in which case we have $\phi=k\neq 0$ along the domain wall.\footnote{In the string theory landscape we expect the scalar field space to be multi-dimensional, in which case the field could come in from the ``side".  It would be interesting to investigate how this affects our results.}

The simplest  potential for $\phi$ which allows it to roll inside our bubble is linear: $V(\phi) = \mu \phi$, which can be thought of as the first term in an expansion around the exact potential on the inflationary slope.  Our first task is to solve for the surfaces of constant $\phi$ in this geometry and with these boundary conditions.

\subsection{Observer Horizons} \label{sec-back}
Before proceeding, it is instructive to determine how much of the reheating surface is inside the past lightcone of an observer at some given time.  We neglect the period of matter domination, which should not strongly affect the conclusions.  We wish to calculate the radius $\xi_r$ of the surface defined by the intersection of the observer's past light cone with reheating, from which we can determine the corresponding values of the coordinates $z$ and $x$ (see \eqref{dstrans} and \eqref{rdtrans}).
The past lightcone of an observer at the origin $\xi=0$ consists of a set of geodesics which in $H_3$ coordinates remain at constant $\theta$ and $\varphi$. In the radiation dominated spacetime the null geodesic is given by%
\beq \label{taudot}
{\partial \tau \over \partial \xi} &=& \sqrt{C \tau} , \\
\Delta \xi &=& {2 \over \sqrt{C} } \Delta (\sqrt{\tau}) .
\eeq%
We want to match this up to the dS space. At reheating we
need to match both the scale factor $a$ and Hubble constant $H$ of the two metrics. This gives%
\beq%
a = \sqrt{C \tau_r} &=& \sinh \tilde \tau_r , \\
{\dot{a} \over a}={1 \over 2 \tau_r}  &=& {1 \over \tanh \tilde \tau_r} ,
\eeq%
where $\tau_r$ is the $H_3$ time of reheating in the radiation dominated spacetime, and $\tilde \tau_r$ is the same for the dS spacetime (we could make the coordinate $\tau$ continuous at the reheating with a shift, but it will be more convenient not to).  We will use tildes for the $H_3$ dS time throughout this subsection; the coordinates on the $H_3$ itself are continuous. The solutions are%
\beq \label{rehvals}%
\tau_r &=& {1 \over 2} \tanh \tilde \tau_r , \nonumber \\
C &=& 2 \sinh \tilde \tau_r \cosh \tilde \tau_r .
\eeq%
Using the fact that temperature redshifts as $1/a$,
\beq%
e^{N_*}\equiv  T_r/T_n = a_n / a_r = \sqrt{\tau_n / \tau_r} ,
\eeq%
where $n$ subscripts refer to the observer's time and this defines $N_*$.

Hence, $\tau_n = e^{2 N_*} \tau_r$. This gives%
\beq%
\xi_r = \frac{(e^{N_*} -1 )}{\cosh \tilde \tau_r}\approx  2 e^{N_* - \tilde \tau_r} = 2 e^{N_*-N},
\eeq%
where $N$ is the number of efolds of inflation and we have used $N, N_* \gg 1$.  In this case $\xi_r \ll 1$ and the flatness problem is solved (because $\xi\sim 1$ is one curvature radius of the $H_3$).  Hence $N_*$ is roughly the number of efolds required to solve the flatness problem.  A simple calculation demonstrates that this same condition is required to solve the horizon problem.

In \cite{wwc} we calculated the location of the intersection of the future lightcone of the collision with the reheating surface, and found that the parameter $t_0$ can be roughly as large as $t_c^3$, where $t_c$ is the time of collision and $t_0$ is the backreaction parameter appearing in \eqref{dsback}.   We found that $t_c$ could be exponentially large and still have the collision within the $\xi=0$ observer's past lightcone. This was done in a toy model by patching a curvature dominated open universe onto the dS space at reheating. One can see from the calculation above that patching on radiation domination rather than flat space dramatically changes this estimate, and that $t_c$ cannot be exponentially large and still be visible to an observer at $\xi=0$.  This justifies taking $t_0 \sim 0$, as we have done here.

Because these collisions break part of the isometry group of open Robertson-Walker cosmology, they define a preferred ``center of mass" frame in which the two bubbles nucleate simultaneously.  In our coordinates, observers at rest in this frame move on geodesics of constant $z$ (or $x$).  The intersection of this set of observers with the set comoving inside the bubble cosmology are geodesics with $z=\xi=0$, spanned by the $H_2$.  In the analysis above we focused on such a ``center of mass" (COM) observer.  Observers who are not at rest in this frame (which we will refer to as ``boosted observers"), including those who formed from matter comoving with respect to the unperturbed part of the reheating surface, will see effects which depend on their boost relative to the COM observers.  Those boosted towards the collision can see bubbles that nucleated farther away in the COM frame (and hence collided with ours later).  For a given scenario, observers boosted towards the collision will see effects enhanced relative to the COM observer (essentially because of Doppler shift).  The opposite applies to observers boosted away.

We will discuss this in more detail in Sections \ref{boosted} and \ref{model}.

\section{Finding The Reheating Surface} \label{sec-surf}
In this section we'll solve for the geometry of the reheating surface. The computation involves several steps. First, we will outline the geometry, causal structure and domain wall trajectory in our bubble. These will determine the boundary conditions on the scalar field. We then solve the scalar field equation for a linear potential with boundary conditions that it is constant on the bubble and domain walls. Armed with the solution for $\phi$ we can find the reheating surface by solving $\phi(z,t) = \phi_0$, where $\phi_0$ is the field value where the field ceases to slowly roll and reheating begins.

\subsection{The Domain Wall Trajectory}
The trajectory of the domain wall in $x,t$ (recall that the wall is homogeneous on the $H_2$) was found in \cite{wwc}%
\beq%
\dot x ^2={1 \over g(t)} \left(-1+{\dot t^2 \over g(t)} \right) ,%
\eeq%
where here a dot denotes differentiation with respect to the proper time $s$ along the wall. At late times the solution for $t(s)$ is%
\beq%
t(s)=T_0 e^{\lambda s} ,%
\eeq%
$T_0$ a constant.  If we assume the other bubble colliding with us is also a dS bubble with radius of curvature $\ell_2$, then%
\beq%
\lambda^2=1+{1\over 4 \kappa^2} \left(\kappa^2 + {1\over \ell_2^2}-1 \right)^2 %
\eeq%
(in units of the radius of curvature of our bubble).
At late times we can use $g(t) \approx t^2$ and plugging in the solution for $t(s)$ we find\footnote{It is straightforward to generalize this to a collision where the other bubble is AdS, results will be similar.}%
\beq%
x(t) = \sqrt{1-\frac{1}{\lambda^2}} \left( {1\over t} - {1\over t_c} \right)+x_c \equiv \alpha \left( {1\over t} - {1\over t_c} \right)+x_c,%
\eeq%
where $(t_c, x_c)$ are the coordinates of the collision.  $0<\alpha<1$ is related to the proper acceleration of the wall, with $\alpha=0$ corresponding to no acceleration and $\alpha=1$ to infinite (so that the trajectory with $\alpha=1$ is null).

\subsection{Null Rays And The Structure Of $H_2$ dS Space}
In this subsection, we'll set up the geometry and boundary conditions we'll need to solve the scalar field equation in the collision background. For dS space, the null rays are given by%
\beq%
{d x \over dt} = \pm {1 \over g} \Rightarrow x(t) = \pm \tan^{-1} t +k ,%
\eeq%
where $k$ is an integration constant. There are three null surfaces important to us:  the two walls of the initial bubble and the lightcone of the collision. We choose to orient our coordinates such that our initial bubble is nucleated at $(t,x)=(0,0)$. Running through each line we find%
\beq \label{null rays}%
\textrm{right wall} &\to& x = \tan^{-1} t, \nonumber \\
\textrm{left wall} &\to& x =- \tan^{-1} t, \\
\textrm{radiation line} &\to& x = \tan^{-1} t  -2\tan^{-1} t_c , \nonumber%
\eeq%
and the collision point is at $(t_c,x_c)=(t_c,-\tan^{-1} t_c)$.

\begin{figure}
\centering \hspace{0.2in}
\includegraphics[width=0.8\textwidth]{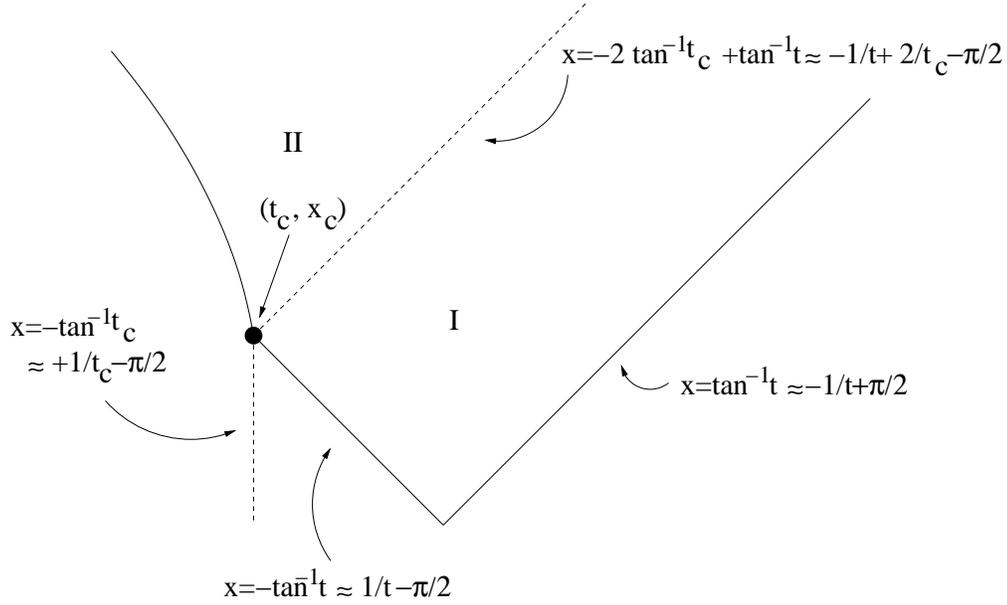} \caption{Sketch of the collision scenario with the null lines labeled and approximate solutions for late times.
\label{dsgeom}}
\end{figure}

We'll also need these rays for large $t$. We can take the asymptotic expansions of \eqref{null rays} to get\footnote{We have here also expanded for large $t_c$, for smaller values of $t_c$ we can convert back by exchanging $1/t_c \leftrightarrow -\tan^{-1} t_c + \pi /2.$}%
\beq%
\textrm{right wall} &\to& x = -{1 \over t} + {\pi \over 2} , \nonumber \\
\textrm{left wall} &\to& x= {1 \over t}  -{\pi \over 2} , \\
\textrm{radiation line} &\to& x = -{1 \over t} + {2\over t_c}-{\pi \over 2} , \nonumber%
\eeq%
and the collision point is at $(t_c,x_c)=(t_c,{1 \over t_c} - {\pi \over 2})$. The geometry and causal structure are summarized in \figref{dsgeom}.  Note that region I is outside the lightcone of the collision, and the geometry can effectively be described in $H_3$ coordinates up to the radiation line.

\subsection{General Solutions For The Scalar Field}
In this subsection we'll find the general solutions to the scalar field equation (with a linear potential) in the regions before and after the collision. Let's begin by looking at region I. Here we can make use of the $H_3$ coordinates. We have%
\beq%
\Box \phi = -{1\over \sinh^3 \tau} \partial_\tau \left( \sinh^3 \tau \partial_\tau \phi(\tau) \right) =  \mu ,%
\eeq%
where $\mu$ is the coefficient of the linear term in the potential.  We require that $\phi$ and its derivative vanish along the bubble walls at $\tau=0$. The general solution satisfying these boundary conditions is%
\beq \label{region1sol}%
\phi(\tau) = {\mu \over 3} \left[\ln{\frac{4 e^\tau}{(1+e^\tau)^2}} -\frac{1}{2}\tanh^2{(\tau/2)} \right] .%
\eeq%
At large and small $\tau$ this becomes%
\beq\label{coordapprox}
\phi(\tau \gg 1) &\approx & -{\mu \over 3} \ln \sinh \tau \approx -{\mu \over 3} \ln (t \cos x) \\
\phi(\tau \ll 1) &\approx & -{\mu \tau^2 \over 8} \approx -{\mu \over 4} \left(\sqrt{1+t^2} \cos x -1\right) ,%
\eeq%
where we have used the coordinate transformations \eqref{dstrans}.

In region II we solve for $\phi$ using the $H_2$ coordinates, as it is in these coordinates that the boundary conditions on the domain wall and collision lightcone are simple. Since the field is constant on the $H_2$ we have%
\beq%
\Box \phi = -{1\over t^2} \partial_t \left[ t^2 (1+t^2) \partial_t \phi\right] +{1\over 1+t^2} \partial^2_x \phi = \mu  .%
\eeq%
The general solution is%
\beq%
\phi(t,x) &=& f(x- \tan^{-1} t)-{1\over t} f'(x-\tan^{-1} t)+g(x+\tan^{-1} t) +{1\over t} g'(x+\tan^{-1} t) \nonumber \\ 
&-& {\mu \over 6} \ln (1+t^2) ,%
\eeq%
where $f$ and $g$ are arbitrary functions and the primes denote derivatives with respect to the argument of the respective function. At large $t$, this is given by%
\beq%
\phi(t,x) \approx f(x+1/t) -{1\over t} f'(x+1/t)+g(x-1/t)+{1\over t} g'(x-1/t) -{\mu \over 3} \ln t .%
\eeq%

\subsection{Matching Solutions}
We have the general solution for the scalar field in the $H_2$ symmetric background, as well as the solution in region I \eqref{coordapprox}, before the collision. To find the particular solution for $\phi$ in region II we make use of two boundary conditions: the field must be continuous at the radiation line, and it must equal a constant at the domain wall, $\phi |_{domain} = k$.  Since reheating occurs after inflation we can use the late time approximations to the general solutions. Carrying out these steps is somewhat technical; the details are in appendix \ref{app-sol}.

\begin{figure}
\centering \hspace{0.2in}
\includegraphics[width=0.8\textwidth]{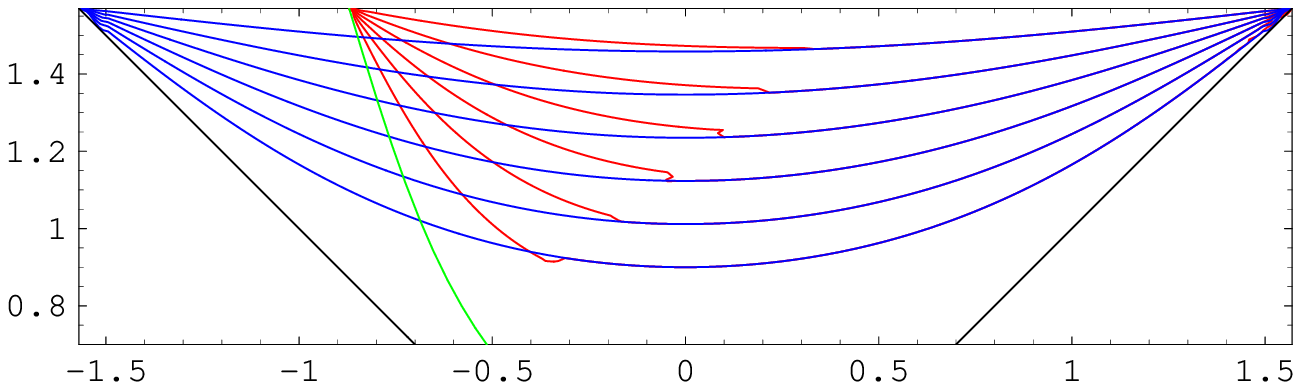} \phantom{xxxxxx}\caption{Plot of surfaces of constant $\phi$ at late times for $N_*=60, \, \alpha=0.3$, $t_c=1$, $\mu=1$, and $k=0$. The blue lines are the surfaces if no collision had occurred, while the red lines are the actual surfaces. Black lines are the bubble walls (projected forward as if there were no collision) and the green line is the domain wall. This plot is in the same coordinates as \figref{coords}.
\label{collision1}}
\end{figure}

In \figref{collision1} we draw an example with $N_*=60, t_c=1, \alpha=0.3, \mu=1$, and $k=0$.  The presence of the domain wall causes the surfaces to ``curl up", eventually going null and then timelike.  That part of the spacetime is very strongly perturbed by the collision---inflation and reheating will not occur there, and the linear techniques used in this paper do not apply.  Since these regions are highly perturbed away from a standard cosmology, the effects would be in drastic conflict with current observations, and thus we will assume that this region is well outside our horizon.

\begin{figure}
\centering \hspace{0.2in}
\includegraphics[width=0.8\textwidth]{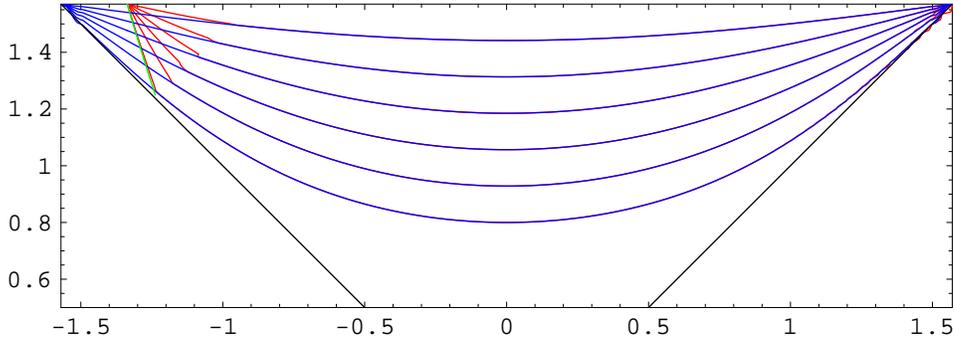} \phantom{xxxxxx} \caption{Plot of surfaces of constant $\phi$ at late times for $N_*=60, \,\alpha=0.3$, $k=0,\, \mu=1$, and $t_c=3$. The blue lines are the surfaces if no collision had occurred, while the red lines are the actual surfaces. Black lines are the bubble walls (projected forward as if there were no collision) and the green line is the domain wall.
\label{collision2}}
\end{figure}

We can make some other general statements. Increasing the collision time $t_c$ pushes the difference between the actual surface and that of no collision farther out in $x$. In \figref{collision2} we draw the same scenario as in \figref{collision1} except with $t_c=3$.  The smaller $\alpha$ is, the more rapidly the surfaces curl up. Taking $\alpha \to 1$ in the case $k=0$ removes the effects of the collision entirely (which is a good check of our numerics).

\begin{figure}
\centering \hspace{0.2in}
\includegraphics[width=.85\textwidth]{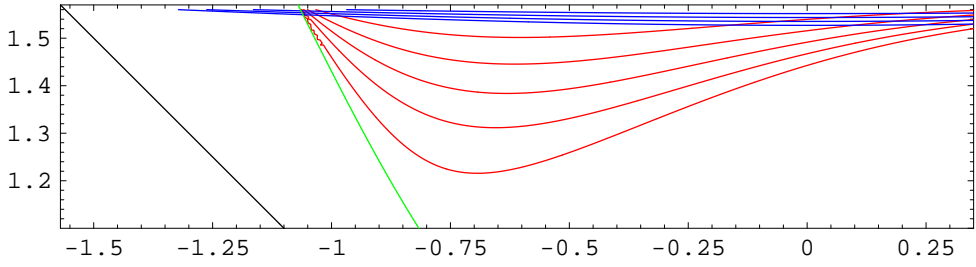} \phantom{xxxxxx}\caption{Plot of surfaces of constant $\phi$ at late times for $N_*=60, \,\alpha=0.5$, $t_c=1$, $\mu=1$, and $k=-1$. The blue lines are the surfaces if no collision had occurred, while the red lines are the actual surfaces. Black lines are the bubble walls (projected forward as if there were no collision) and the green line is the domain wall.
\label{redcoll}}
\end{figure}

For $k<0$ we find that the reheating surface bends down relative to the surface of the no collision scenario. In \figref{redcoll} we draw a case with $N_*=60,t_c=1, \alpha=0.5, k=-1$, and $\mu=1$.  Increasing $t_c$ causes less of the sky to be covered by the collision, while the smaller $\alpha$ gets the more pronounced and closer to the observer the effects are. Taking $k>0$ is similar to the scenarios with $k=0$ except that the repulsion from the domain wall will be increased.

\section{The Redshift} \label{sec-red}
In this section we will use our solution  for the modified reheating surface to find the redshift from it to observers in the bubble. Using the results of \secref{sec-back}, we find the projection of the observer's past lightcone on the reheating surface and calculate the redshift as a function of angle, taking into account the doppler shift due to the fact that the reheating surface is not comoving.  We will again begin by considering COM ($z=0$) observers, and later generalize to boosted observers.  In doing this analysis we will ignore the effects of gravitational backreaction from the collision and modified reheating surface, as well as the modification to $a(\tau)$ from spatial curvature .  These approximations are justified when $N \gg 1$, at least for observers away from the part of the spacetime that is strongly modified.

\subsection{Null Geodesics In Radiation Dominated Spacetime}

Starting from the results of \secref{sec-back}, we can express the geodesics in $H_2$ coordinates using the coordinate transformations \eqref{rdtrans}%
\beq%
t &=& \left( {C \over 2} (\xi_r - \xi) + \sqrt{C \tau_r} \right) \sqrt{\cosh^2 \xi-\sinh^2 \xi \sin^2\theta} , \nonumber \\
z &=& \left( {C \over 2} (\xi_r - \xi) + \sqrt{C \tau_r} \right) \sinh \xi \cos \theta , \\
\tanh \rho &=& \tanh \xi \sin \theta ,\nonumber%
\eeq%
where for each geodesic $\theta$ is a constant and the set of all geodesics is given by $\theta$ taking all values from $0$ to $\pi$, with 
\beq%
C \approx {1\over 2} e^{2N} , \quad \tau_r \approx {1\over 2} . %
\eeq%
Since $z_n=0$ the observer's time $t_n$ is given by%
\beq%
t_n =\sqrt{C \tau_n} \approx \sqrt{ {1 \over 4} e^{2 N} e^{2 N_*} } \approx {1\over 2} e^{N+N_*} .%
\eeq%
Using this we can solve for $\xi_r$ by setting $t(\xi=0) = t_n$, giving%
\beq%
\xi_r = 4 e^{-2 N} \left(t_f - {1\over 2} e^N \right) \approx 2 e^{N_* -N} .%
\eeq%
Putting this all together we get%
\beq%
t &\approx& {1 \over 2} \left ( e^{N_*+N} + e^N-{\xi\over 2} e^{2N} \right) \sqrt{\cosh^2 \xi-\sinh^2 \xi \sin^2\theta} , \nonumber \\
z &\approx& {1 \over 2} \left ( e^{N_*+N} + e^N-{\xi\over 2} e^{2N} \right)  \sinh \xi \cos \theta .%
\eeq%

To compute the redshift we have to take into account the relative motion of the reheating surface and observer.  To do so we need another piece of information about the geodesics, namely their 4-momenta at both the time of emission and when they intersect the observer. We can again make use of the $H_3$ coordinates. Differentiating both sides of the $z$ equation of \eqref{rdtrans} with respect to the affine parameter we find%
\beq%
\dot z =p^z= \sqrt{C \tau} \cos \theta \left( {1 \over 2 \tau} \sinh \xi \, p^\tau+ \cosh \xi \, p^\xi \right) ,%
\eeq%
where $p^\tau = \dot \tau$ and $p^\xi = \dot \xi$.
Since these geodesics have constant $\theta$, $p_\xi = g_{\xi \xi} p^\xi$ is conserved. From \eqref{taudot}, $p^\tau = - \sqrt{C \tau} \, p^\xi$. We can solve for $p^t$ in a similar way to get%
\beq%
p^z &=& -{C z \over 2 (t^2-z^2)^{3/2}} \, p_\xi+  {t \cosh \rho \over t^2-z^2} \, p_\xi \cos\theta, \nonumber \\[-.3cm]
\\[-.3cm]
p^t &=& -{C t \over 2 (t^2-z^2)^{3/2}} \, p_\xi + {z \cosh \rho \over t^2-z^2} \, p_\xi \cos\theta \nonumber .
\eeq%

\subsection{Boosted Observers}\label{boosted}

So far we have considered the special case of observers at rest in the frame in which the bubbles nucleated simultaneously.  In our coordinates these COM observers are those in the $z=0$ plane, and can always be taken to be at the origin of the $H_2$ ($\rho=0$) by a shift in the origin of coordinates.   

An observer comoving in the open universe inside the bubble is not in general at rest in this frame.  But since the boosts ``parallel" to the domain wall are trivial ({\it i.e.} one can always choose coordinates so any comoving geodesic lies in the $t,z$ plane and at the origin of the $H_2$), the most general comoving observer is related to an observer at $z=0$, due to the $H_3$ isometry, by a ``boost" in the $z$ direction only.  The isometry is
\beq
\tau' &=& \tau \\  \nonumber
\cosh \xi' &=& \gamma \left( \cosh \xi - \beta \sinh \xi \cos \theta \right) \\  \nonumber
\sinh \xi' \cos \theta' &=& \gamma \left( \sinh \xi \cos \theta - \beta \cosh \xi \right), 
\eeq
where $\gamma = (1-\beta^2)^{-1/2}$ is a ``boost" around the nucleation point of our bubble.  $\beta > 0$ corresponds to an observer closer to the collision (boosted to the left in our diagrams).  In $H_2$ coordinates this becomes
\beq
t' \cosh \rho' &=& \gamma \left( t \cosh \rho - \beta z \right) \\  \nonumber
z' &=& \gamma \left( z - \beta t \cosh \rho \right) \\  \nonumber
t'^2 &=& t^2 - z^2 + z'^2.
\eeq

It is clear that a boosted observer will not see the same redshift pattern from the collision in the CMB as a COM observer, even at equal comoving time.  To compute this effect we need to repeat the calculation we have just done, only now for an observer following the boosted trajectory.  We can find the boosted lightcone by acting on the lightcone of the $z=0$ observer with the boost (this works since the boost is an isometry of the unperturbed spacetime).  For brevity we will not include the details here; the calculation is straightforward.

One should be careful to distinguish two cases:  structures whose comoving trajectories intersect the reheating surface in the region outside the lightcone of the collision and only pass into it much later, versus those that do not.  We expect the first class of objects to be comoving in the unperturbed spacetime (up to peculiar velocities), but the second class formed in a part of the spacetime where the reheating surface was perturbed.  These structures may be locally at rest with respect to the frame defined by the modified reheating surface, and will possess a coherent ``peculiar velocity" with respect to the first class.  Our analysis here is valid only for observers in the first class, but in the regions of the space where the effects are small it should be approximately correct.

\subsection{Finding The Redshift}
Armed with the geodesics and the reheating surface, we have all the information we need to compute the redshift as a function of the angle on an observer's sky.  For a given angle $\theta, \varphi$ on the observers sky, we follow the null geodesic back to where it intersects the modified reheating surface.  We take the dot product of the geodesic's 4-momentum with the unit normal to the reheating surface at the intersection point, and then the dot product with the tangent vector to the observer's geodesic.  The ratio of these two quantities gives us the redshift in the approximation we mentioned before. 

We first calculate the reheating surface and normals from our solution for $\phi$ in \secref{sec-surf}. The surface is found by solving $\phi(t,x) = \phi_0$ for $t$ as a function of $x$ and $\phi_0$.   For a given potential $V(\phi)$, $\phi_0$ determines the number of inflationary e-folds $N$.   In the end we have the modified reheating surface for a given set of parameters $(N,N_*,t_c,\alpha,\mu,k)$.

\begin{figure}
\centering \hspace{0.2in}
\includegraphics[width=0.5\textwidth]{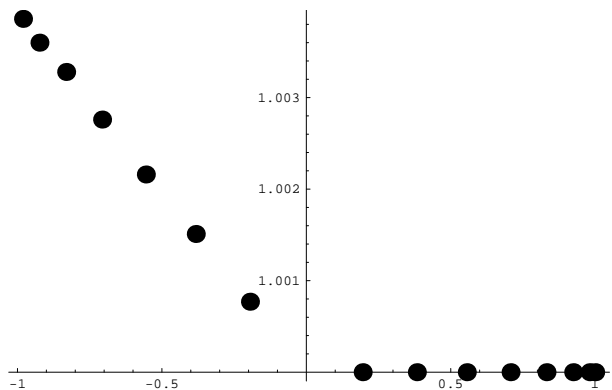} \caption{Relative redshift ($r_s/r_0$) versus $\cos \theta$ for the scenario ($t_c=1, \alpha=0.5, N-N_*=5, k=0, \mu=1$). Relative redshift greater than one indicates higher relative temperature. 
\label{redshift}}
\end{figure}

Once we have the intersection point, we can compute the vectors normal to the surface by taking the exterior derivative of $\phi(t,z)$ (in the RD part of the spacetime). We normalize these using the RD metric \eqref{rdmetric}. The tangent 1-form along the observer's trajectory is proportional to $dt$ for an unboosted observer at $z=0$, and to $d \tau \sim t dt - z dz$ in general. The redshift is given by%
\beq%
r_s \equiv {\nu_n \over \nu_r} = {E_n \over E_r} = {-(p ^\mu n _\mu)_n \over -(p^\mu n _\mu)_r} ,%
\eeq%
where $n$ refers to the observer's position, $r$ to the point  on the reheating surface, and $n_\mu$ are the unit 1-forms at each. Recall if there was no collision then the redshift would be%
\beq%
r_0 = e^{-N_*} ,%
\eeq%
which is also the redshift for any geodesics that originate from parts of the reheating surface that lie outside the future lightcone of the collision. In \figref{redshift} we show some results for the scenario ($t_c=1, \alpha=0.5, N-N_*=5, k=0, \mu=1$). We plot $r_s / r_0$ as a function of $\cos\theta$ where $0<\theta<\pi$ is the angle on the sky with $\theta =\pi$ pointing towards the collision (the redshifts are axially symmetric). 

We can make some general observations about the redshift. The modified reheating surface leads to bluer photons when $k \geq 0$, the bluest coming from the angle closest to the domain wall. This makes sense, as the reheating surface is closer to the observer than it would be without the collision.   However, the doppler shift from the motion of the reheating surface with respect to the observer also plays a role. The effect is primarily a dipole, but only on the range of angles for which null geodesics intersect the modified reheating surface. Once we are viewing angles which correspond to parts of the reheating surface outside the future lightcone of the collision, the redshift is $r_0$. The smaller $\alpha$ is, the more pronounced the effects of the collision are, as the domain wall is closer to the observer and accelerating away more slowly. Raising $t_c$ means less of the sky is within the future lightcone of the collision.  Thus, there are a wide range of values for both the maximum blueshift and the angular size of the part of the sky affected by the dipole piece. If $t_c$ is much larger than $1$ in Hubble units the COM  observer is likely not within the future lightcone of the collision (as we mentioned in \secref{sec-back}). 
 
\begin{figure}
\centering \hspace{0.2in}
\includegraphics[width=0.5\textwidth]{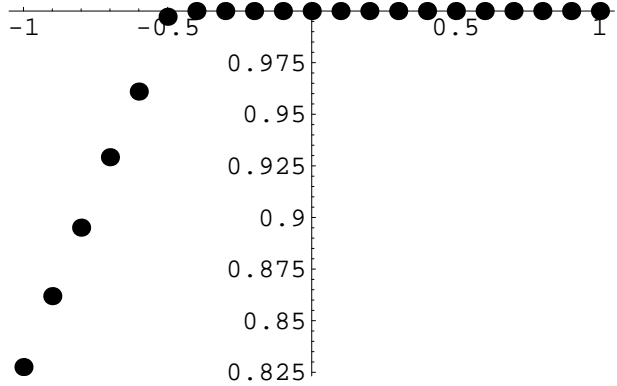} \caption{Relative redshift ($r_s/r_0$) versus $\cos \theta$ for the scenario ($t_c=1.05, \alpha=0.1, N-N_*=3, k=-1, \mu=1$). 
\label{reddershift}}
\end{figure}

If we look at scenarios where $k<0$ we see that the modified reheating surface will lead to slightly redder photons, again the reddest coming from the angle closest to the collision direction. This is due to the reheating occurring earlier in that direction. In \figref{reddershift} we display a typical scenario with ($t_c=1.05, \alpha=0.1, N-N_*=3, k=-1, \mu=1$). 

\begin{figure}
\centering \hspace{0.2in}
\includegraphics[width=0.5\textwidth]{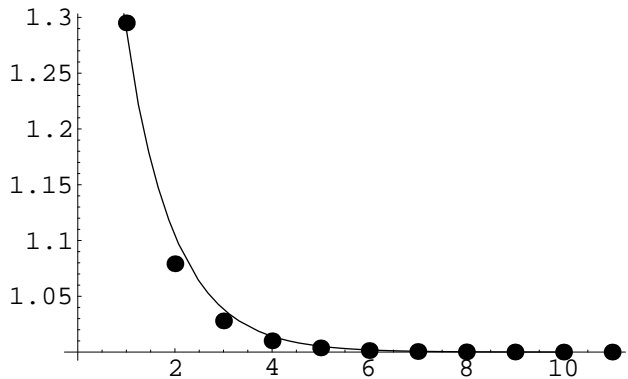} \caption{Relative redshift ($r_s/r_0$) versus $N-N_*$ for the scenario ($t_c=1, \alpha=0.5, k=0, \mu=1)$ at angle $\cos \theta=-0.9$. The dots are actual values and the line is a best fit line with equation $r_s/r_0 =1+0.77 \,e^{N_*-N}$.
\label{nfitting}}
\end{figure}

Let's see how this effect varies with the number of e-folds $N$. In \figref{nfitting} we show some sample values for a collision with $t_c=1, \, \alpha=0.5, k=0, \mu=1$ at angle $\cos \theta=-0.9$. We plot it for $1<N-N_*<11$ and draw a best fit line. The best fit line at this angle is $r_s/r_0 \approx 1 + 0.77 \, e^{N_*-N}$, which decreases exponentially in $N$ as expected from inflation. We expect that the general behavior will go as%
\beq%
{r_s \over r_0} (\alpha, t_c, \theta,k,\mu) \approx 1 + \xi(\alpha, t_c,\theta,k,\mu) e^{N_*-N} ,%
\eeq%
where $0<|\xi|<1$ is a function of the specifics of the collision and the angle. Thus, unboosted observers can detect effects from the collision so long as the number of e-folds of inflation is not too large.

Observers boosted towards the collision will see more of the modified part of the space, and hence the angular size of the affected spot will be larger.  On the other hand, observers boosted away see the opposite (smaller, weaker spots).  For both, the angular dependence within the spot is similar in form to the COM observer, constant away from the collision and linear towards the collision.  It is interesting to note that even long inflation cannot completely remove the effects on the skies of some boosted observers, since there are always parts of the space near the domain wall.  However these observers are not well-approximated by the treatment carried out here, since in the limit of large boost and long inflation they formed from strongly affected parts of the reheating surface.  Nevertheless, it is possible that such observers see a more or less standard cosmology.  The study of this question is beyond the reach of this paper.

$t_c$ larger than of order 1 will take the COM observer outside the future lightcone of the collision.  However, it is again true that a sufficiently large boost towards the collision will bring the observer back into the collision's lightcone.  Hence, no matter what the values of $t_c$ and $N$ are, some set of observers will always be able to see the effects of the collision.  How probable we are to find ourselves in that set is again beyond the scope of this paper, but see \cite{futurematt2}.

\section{A Simple Model}\label{model}
The preceding results show that the most natural way to describe the signature of a collision is in position space, with a simple piecewise linear behavior for the redshift as a function of  $\cos\, \theta$.
However, in order to make direct comparisons to previous CMB analyses,  we will use this result to explore the consequences for the CMB multipoles.  
Looking at Figures \ref{redshift} and \ref{reddershift} we see that the redshift can be approximately modeled as%
\beq%
r(\n) = r_0 \left( \Theta(x_T-x) { (M-1) \over x_T+1} (x_T-x)+1 \right) ,
\eeq%
where $r_0$ is a constant that gives the redshift with no collision, and $x=\cos \theta$. The parameter $M$ gives the maximum value of the relative blue or redshift due to the collision. The parameter $x_T$ is the cosine of the angle where the forward light cone of the collision ends, i.e. for $x > x_T$ we are looking at the unperturbed reheating surface.
\begin{figure}
\centering \hspace{0.2in}
\includegraphics[width=0.5\textwidth]{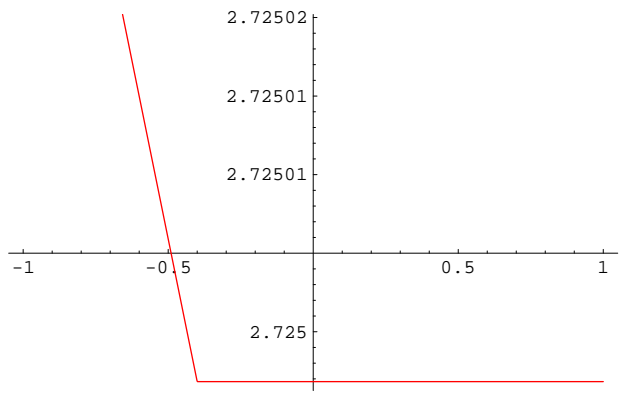}\phantom{xxxx} \caption{An example of the temperature (excluding fluctuations) as a function of angle in our toy model. 
\label{toymodel}}
\end{figure}

Including the density fluctuations at reheating, the temperature today is%
\beq%
T(\n) = T_0 ' r(\n) (1 + \delta (\n) ) ,%
\eeq%
where $T_0 '$ is a constant that sets the overall temperature and $\delta (\n)$ are the temperature anisotropies from inflationary density perturbations. We will take $\delta(\n)$ to be Gaussian random variables with zero mean.  We plot an example of the function $T_0' r(\n)$ in \figref{toymodel}.  

To find the CMB multipoles we need the fluctuations of the temperature about its average value. This is given by%
\beq%
{\delta T(\n) \over T_0 } = { T_0 ' \, r(\n) (1 + \delta(\n) ) - T_0 \over T_0} ,%
\eeq%
where%
\beq%
T_0 = {T_0' \over 4 \pi} \int d^2 \n \, r(\n) %
\eeq%
is the average temperature and we've used the fact that $\int d^2 \n ~\delta(\n)$ = 0. 

Several authors have considered anisotropic CMB spectra where the temperature fluctuations are modulated by a function \cite{hu,Eriksen:2007pc,sean1}. We can parallel some of the formalism in \cite{hu} with some modifications. We write%
\beq%
{\delta T(\n) \over T_0 } = A(\n) + f (1 + w(\n) ) B(\n) ,%
\eeq%
where $B(\n) = \delta (\n)$, $f=T_0 ' / T_0$ and%
\beq%
w(\n) = \Theta(x_T-x) (x_T-x) { (M-1) \over x_T+1} .%
\eeq%
In \cite{hu}, $A(\n)$  is a random variable but with a distribution that differs from that of $\delta(\n)$. 
In our case, we want to find the expected CMB multipoles with fixed parameters for the collision; {\it i.e.}, we will average over density fluctuations while holding fixed the collision redshift function $r(\n)$.  Ideally, we would use a microscopic model (for example, the string theory landscape) to compute the probability distribution for the parameters $t_c$ and $\beta$---which are determined by the location and time of nucleation of the other bubble and are random, since the nucleation is a quantum event---and hence a probability distribution for $r(\n)$.    However, without detailed knowledge of the probability measure for collisions we are not currently equipped to carry out this procedure.  In fact, there could be multiple collisions in the past lightcone of the observer, each with an associated time and position for its nucleation.  It would interesting to study this, but here we will proceed to study a single collision with fixed values for the parameters.

$A(\n)$ is given by
\beq%
A(\n) = { T_0 ' (w (\n) +1 ) \over T_0} -1 .%
\eeq%
We can expand the fluctuations in multipoles via spherical harmonics. 
\beq%
{\delta T(\n) \over T_0 } = \sum_{l,m} t_{lm} Y_{lm} (\n) .%
\eeq%
For details on our conventions for the spherical harmonics see appendix \ref{app-ylm}. We can expand $A,B$ and $w$ in multipoles as well%
\beq%
A(\n) &=& \sum_{lm} a_{lm} Y_{lm} (\n) , \nonumber \\
B(\n) &=& \sum_{lm} b_{lm} Y_{lm} (\n) ,  \\
w(\n) &=& \sum_{lm} w_{lm} Y_{lm} (\n) \nonumber.%
\eeq%
Since we have chosen coordinates such that $\theta=\pi$ corresponds to the collision direction, both $w$ and $A$ have only $m=0$ multipoles.  Other coordinates can be obtained via a straightforward rotation using Euler angles.

We are now in a position to write the multipoles of the temperature fluctuations. The coupling between $w(\n)$ and $B(\n)$ becomes a convolution in fourier space%
\beq \label{tlm}%
t_{lm} = a_{l0} \delta_{m0} + f b_{lm} + f \sum_{l_1 l_2} R_{lm}^{l_1 l_2} b_{l_2 m} ,%
\eeq%
where the coupling matrix is written in terms of Wigner 3$j$ symbols%
\beq%
R_{lm}^{l_1 l_2} = (-1)^m  \sqrt{ { (2l+1)(2l_1+1)(2l_2+1) \over 4 \pi} }  \left( \begin{array}{ccc} l_1 & l_2 & l \\
0 & 0 & 0 \\ \end{array} \right) \left( \begin{array}{ccc} l_1 & l_2 & l \\
0 & m & -m \\ \end{array} \right)  w_{l_1} ,%
\eeq%
where $w_l = w_{l0}$. 

The functions $A(\n)$ and $w(\n)$ have non-zero 1-point functions:
\beq%
\langle a_{lm} \rangle &=& a_{lm} , \nonumber \\[-.3cm]
\\[-.3cm]
\langle w_{l} \rangle &=& w_l \nonumber,%
\eeq%
while%
\beq%
\langle b_{lm} \rangle = 0 .%
\eeq%
Thus we can quickly compute the multipoles for the one-point function%
\beq%
\langle t_{lm} \rangle = \langle a_{lm} \rangle = a_{lm} \delta_{m0} .%
\eeq%
The only anisotropy we see at this level is a result of the collision itself.  Consistency with the observed CMB dipole requires that $|M-1|\lesssim 10^{-3}$.

A more useful object is the 2-point function. We have%
\beq%
\langle a_{lm} ^*  a_{l' m} \rangle &=& \delta_{m0}  a^* _{l0} a_{l'0} ,\nonumber \\
\langle b_{lm} ^* b_{l' m} \rangle &=& \delta_{l l'} C_l ^{bb} ,  \\
\langle a_{lm} ^* b_{l' m} \rangle &=& a_{lm} ^* \langle b_{l'm} \rangle =0 \nonumber.%
\eeq%
$C_l ^{bb}$ is the two point function in the absence of a collision. We'll discuss this in a bit.

Squaring \eqref{tlm} and taking ensemble averages using the above rules we get%
\beq%
\langle t^* _{lm} t_{l'm} \rangle = \delta_{m0}  a^* _{l0} a_{l'0} +f^2 \delta_{ll'} C_l^{bb} +f^2 \sum_{l_1} R_{lm}^{l_1l'} C_{l'}^{bb} + ( l \leftrightarrow l') + f^2 \sum_{l l_2 l_1'} R_{lm}^{l_1 l_2} R_{l'm}^{l_1' l_2} C_{l_2} ^{bb}.\quad %
\eeq%
This determines the two-point function in terms of $C_l ^{bb}$ and the collision redshift function. As a first approximation we can assume that $C_l ^{bb}$ comes from primordial fluctuations that are unaffected by the collision.  We use a spectrum generated with CMBFAST code \cite{cmbfast1,cmbfast2} using concordance cosmology values from WMAP  \cite{wmap,wmapsky}.

\subsection{Results}
\begin{figure}
\centering \hspace{0.2in}
\includegraphics[width=0.5\textwidth]{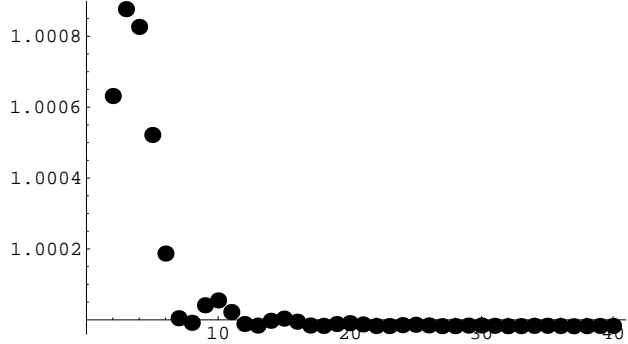} \caption{$C_l / C_l ^{(0)}$ vs. $l$ for the scenario $M-1=2 \times 10^{-5}$, $x_T=0.8$
\label{cmbmodel1}}
\end{figure}

In this section we will present our results for a few simulations of our toy model. The effects are predominantly on large angular scales, so we focus on the lowest set of $l$ modes. We will look at two examples, one where the collision takes up most of the sky, and another where it covers only a small fraction. For each scenario we present two figures. One will display the power in each $l$ mode compared to a scenario where no collision occurs, the other the power asymmetry as a function of $l$ between the two hemispheres defined by the collision direction.

The $C_l$'s are%
\beq%
C_l = {1 \over (2l+1)} \sum_{m=-l}^{m=l} \langle t^*_{lm} t_{lm} \rangle .%
\eeq%
Due to the fact that the $a_{l m}$'s only contribute to the $m=0$ mode of each multipole, these terms gain more power from the collision than the modes with $m \neq 0$ do.  This is due to the fact that the effects of the collision are azimuthally symmetric.

We compute the power asymmetry between the left and right hemispheres using a crude version of the Gabor transform method \cite{gabor} with a top hat window function centered in either hemisphere. There will be effects due to ringing from the edges of the top hat, but these don't greatly affect the lower multipoles.

\begin{figure}
\centering \hspace{0.2in}
\includegraphics[width=0.5\textwidth]{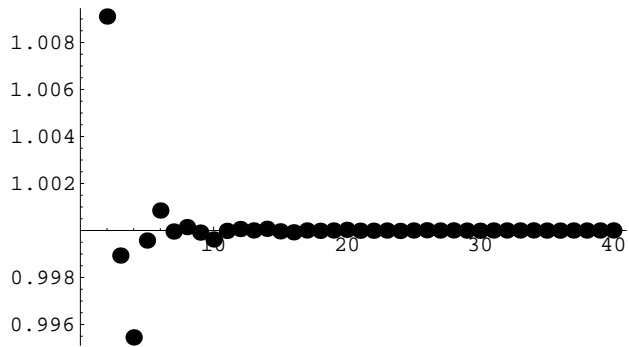} \caption{$C_l ^{left} / C_l ^{right}$ vs. $l$ for the scenario $M-1=2 \times 10^{-5}$, $x_T=0.8$
\label{model1asy}}
\end{figure}

For our first example, we choose parameters $M-1=2 \times 10^{-5}$ and $x_T=0.8$. The angular radius of the collision is $143$ degrees. Note that the total temperature difference between the two poles is well below the dipole due to the Earth's peculiar motion. In \figref{cmbmodel1} we plot $C_l / C_l ^{(0)}$ for this scenario. As expected, since the collision takes up a large portion of the sky, the lowest multipoles are the most affected. In \figref{model1asy} we show the hemispherical power asymmetry.

In our second scenario the collision takes up only a small part of the sky, and the collision causes a redshift within the disk. We choose $M-1=-7.3 \times 10^{-6}$ and $x_T=-0.99$, giving an angular radius of about $\theta_T= 180^\circ - \cos^{-1} x_T = 7^\circ $ . With these parameters our sky has a cold spot, with the coldest point about 20$\mu$K cooler than the average and the effect falling off with radius from the center.  We have chosen the angular radius and the temperature profile to approximate the WMAP cold spot smoothed over a scale of $\sim 5^\circ$ \cite{coldspot1,coldspot2,coldspot3,coldspot4,coldspot5}.

\begin{figure}
\centering \hspace{0.2in}
\includegraphics[width=0.5\textwidth]{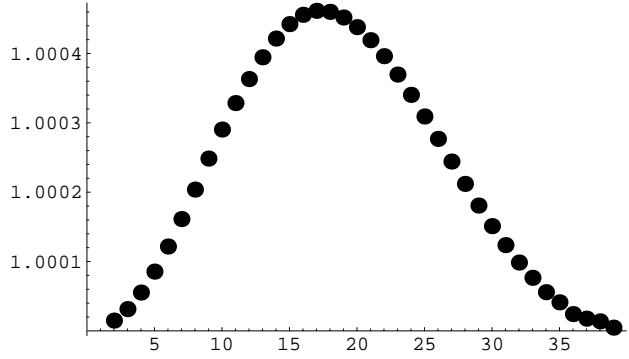} \caption{$C_l / C_l ^{(0)}$ vs. $l$ for the scenario $M-1=-7.3 \times 10^{-6}$, $x_T=-.99$.  
\label{cmbmodel2}}
\end{figure}

In \figref{cmbmodel2} one can see that the power is shifted slightly towards higher multipoles, with a first peak in excess power around $l=17$. There will be a smaller secondary peak at the first harmonic. The quadrapole receives only a small boost in power, reducing it relative to nearby multipoles.

\begin{figure}
\centering \hspace{0.2in}
\includegraphics[width=0.5\textwidth]{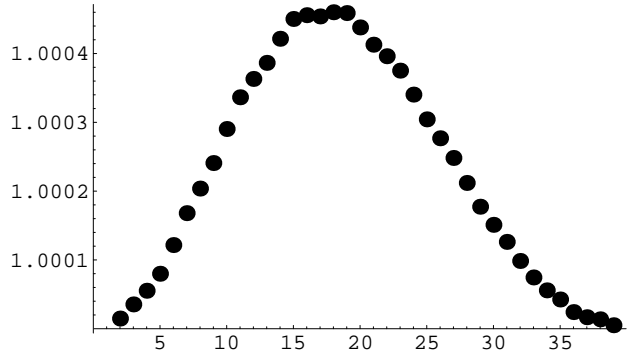} \caption{$C_l ^{left} / C_l ^{right}$ vs. $l$ for the scenario $M-1=-7.3 \times 10^{-6}$, $x_T=-.99$.
\label{model2asy}}
\end{figure}

In \figref{model2asy} we show the hemispherical power asymmetry, which is nearly identical to the previous figure (since the effects of the collision are now localized wholly in the left hemisphere). We see that collisions of this type can lead to measurable hemispherical power asymmetries.

We have presented results for two specific choices of parameters.  The results for general parameters all have certain features in common:  most of the excess power is in the $m=0$ modes, the temperature function is monotonic with angle within the affected disk, the power spectra are affected primarily in the $l$ modes corresponding to the size of the disk and its harmonics, and there is a hemispherical power asymmetry with a magnitude that depends on the size and intensity of the disk.

\paragraph{Non-Gaussianities:}
The effects of the bubble collision induce sizable non-Gaussianities in the CMB temperature map.  However, at least for the two examples presented here the non-Gaussianity is significant only at low $l$.  In particular, $f^{\rm local}_{NL}$ is very small, and $f^{\rm equilateral}_{NL}$ is strongly dependent on $l$:  it oscillates in sign, reaches a maximum near the $l$ corresponding to the disk size, and damps rapidly (with an envelope similar to the plots of the excess power in the 2-point function).  We suspect that an analysis of the WMAP data would not constrain these models significantly.  We will mention a better statistical test in the conclusions. 

\section{Conclusions}
In this paper we have computed the effects of a cosmological bubble collision on the cosmic microwave background, using a variety of approximations.  Previously we focused on a range of parameters where gravitational backreaction significantly affects the CMB even outside the collision disk \cite{wwc}.  However this effect is probably too weak to be observable in a more realistic model.  Instead, in this paper we concentrate on what we expect to be a stronger effect which arises from the fact that the reheating surface---defined by a constant value of the tunneling scalar field---is perturbed from the case with no collision.  Because of this the redshift back to the last scattering surface depends in a distinct and specific way on the angle from the collision direction (but not on the azimuthal angle).  In particular, our calculations demonstrate that this has the behavior seen in \figref{toymodel}---no effect for $\cos \theta > \cos (\pi -\theta_T)$  and linear in $\cos \theta$ for $\cos \theta < \cos (\pi-\theta_T)$ (where $\theta_T$ is the angular radius of the affected disk and depends on the collision parameters).   This form of anisotropy is seen for all observers in the part of the spacetime where our approximations are valid, but different observers will measure different values for $\theta_T$ and the temperature gradient inside the disk.

To compare to the CMB maps, we assume that the inflationary perturbations are unaffected by the collision and calculate the temperature perturbation map as well as the two point function for fluctuations.  The deviations in the  $C_l$'s are primarily at large scales (low $l$), and depending on the transition angle, affect $l$'s at harmonics of $\pi/\theta_T$.  Non-Gaussianities can be relatively large depending on the parameters, and are best detected using equilateral triangles.

It is possible that this scenario can explain some of the current cosmological anomalies, but the analysis we have done so far is not detailed enough to provide a ``smoking gun".  A good statistic for the effect described in this paper would be to pick an origin, measure the average CMB temperature in a disk of some angular size around that origin, and then find the point and angular radius which maximizes the effect.    We have not performed this analysis, but the results would be very interesting.

Many open questions remain.  It should be possible to calculate how the inflationary perturbations are affected by the collision and include this effect.  Aside from the CMB, there should be anisotropies in other cosmological observables, such as large scale structure.  As we mentioned in the introduction, one expects coherent peculiar velocities due to the fact that the affected part of the reheating surface is not comoving in the frame of the unaffected region, as observed in \cite{darkflow}.  Other potential signals could be detected using 21 cm observations, CMB polarization \cite{Dvorkin:2007jp}, and better precision on measurements of spatial curvature. Finally, it is important to estimate how many bubble collisions to expect in our past lightcone, and what the size distribution of disks should be on the sky.

\section*{Acknowledgements}
We thank Hans Kristian Eriksen, Ben Freivogel, Lam Hui, Eugene Lim,  Alberto Nicolis,  Stephen Shenker, Lenny Susskind, and especially Marc Manera, Ignacy Sawicki, and Roman Scoccimarro for valuable discussions. TSL was supported in part by NSF CAREER grant PHY-0645435 and in part by NSF grants PHY-0245068.  SC was supported in part by NSF CAREER grant PHY-0449818 and DOE grant \# DE-FG02-06ER41417.  MK is supported by NSF CAREER grant PHY-0645435.

\appendix
\section{Solving For The Scalar Field} \label{app-sol}
In this appendix we will describe how to solve for the scalar field in the collision geometry. In region I we already have the solution \eqref{region1sol}. In region II we have to match our solution to the region I solution at the radiation line%
\beq%
\phi_{II} (t,x=2/t_c-1/t-\pi/2) &=& f(2/t_c-\pi/2)-{1\over t} f'(2/t_c-\pi/2)+g(-2/t+2/t_c-\pi/2) \nonumber \\
&+& {1\over t} g'(-2/t+2/t_c-\pi/2)-{\mu \over 3} \ln t \nonumber \\[-.3cm]
\\[-.3cm] 
&=& -{\mu \over 3} \ln t -{\mu\over3} \ln \left[ \cos \left(-1/t+2/t_c-\pi/2\right)\right] \nonumber \\
&=& \phi_{I} (t,x=2/t_c-1/t-\pi/2) \nonumber.%
\eeq%
This is a differential equation for $g$. We can write $u \equiv -2/t+2/t_c-\pi/2$ and get%
\beq%
g(u)+\left({1\over t_c}-{\pi \over 4} - {u \over 2} \right) g'(u) &=& -{\mu \over 3} \ln\left[\cos \left(u/2+1/t_c-\pi/4\right)\right] \nonumber \\[-.3cm]
\\[-.3cm]
&-& f(2/t_c-\pi/2)+\left({1 \over t_c}-{\pi \over 4}-{u\over 2} \right) f'(2/t_c-\pi/2) \nonumber.%
\eeq%
This equation is difficult to solve due to the $\ln \cos$ term. We can use the fact that $t\gg 1$ and expand this term about $u\approx 2/t_c-\pi/2$. We find%
\beq%
g(u) &+& \left({1\over t_c}-{\pi \over 4} - {u \over 2} \right) g'(u) \approx -{\mu \over 3} \ln\left[\sin (2/t_c)\right] -{\mu\over 6} \cot(2/t_c) \left(u-{2\over t_c}+{\pi \over 2}\right)\nonumber \\[-.3cm]
\\[-.3cm]
&-& f(2/t_c-\pi/2)+\left({1 \over t_c}-{\pi \over 4}-{u\over 2} \right) f'(2/t_c-\pi/2) \nonumber.
\eeq%
The solution is%
\beq \label{gsoln}%
g(u) &\approx& -f(2/t_c-\pi/2)-{\mu\over 3} \ln \sin(2/t_c) \nonumber \\[-.3cm]
\\[-.3cm]
&-& \left(u-{2\over t_c}+{\pi \over 2} \right) \left( f'(2/t_c-\pi/2) +{\mu\over 3} \cot(2/t_c) \right) +A \left(u-{2\over t_c}+{\pi \over 2} \right)^2 \nonumber,%
\eeq%
where $A$ is an integration constant we will determine later. 

At the domain wall we have%
\beq%
x-{1\over t} &=& (\alpha-1) \left({1\over t} - {1\over t_c} \right) -{\pi \over 2} , \\
x+{1\over t} &=& {\alpha+1 \over t}-{\alpha-1 \over t_c} -{\pi \over 2} .%
\eeq%
The condition for the field to be constant along the domain wall becomes%
\beq%
\phi_{II}(t, x=\alpha(1/t-1/t_c)  + 1/t_c -\pi/2 ~) &=& \nonumber \\ f\left( {\alpha+1 \over t}-{\alpha-1 \over t_c} -{\pi \over 2}\right)
&-&{1\over t} f'\left({\alpha+1 \over t}-{\alpha-1 \over t_c} -{\pi \over 2}\right)\nonumber\\[-.2cm]
\\[-.2cm]  
+ g\left((\alpha-1) \left({1\over t} - {1\over t_c} \right) -{\pi \over 2} \right)
&+&{1\over t} g'\left((\alpha-1) \left({1\over t} - {1\over t_c} \right) -{\pi \over 2}\right) \nonumber \\
&-& {\mu \over 3} \ln t =k \nonumber,%
\eeq%
where $g$ is given by \eqref{gsoln} and $k$ is the value of the field along the wall. This equation is difficult to solve in closed form, but can be solved using Mathematica. We follow a similar procedure as we did to obtain the solution for $g$. We rewrite everything so that the argument inside $f$ becomes a dummy variable $u$. In this case, $u={\alpha+1 \over t}-{\alpha-1 \over t_c} -{\pi \over 2}$. Then we re-express all $t$ dependence in terms of $u$. This gives a differential equation for $f(u)$ once we utilize \eqref{gsoln} for the correct argument inside of $g$ as a function of something involving $u$. Then one solves the equation for $f(u)$. 

The general solution will have two integration constants, one each from the differential equations for $f$ and $g$. Demanding that $f(2/t_c -\pi/2)$ is continuous at the radiation line fixes both constants to lowest order in $1/t$:  
\beq%
\phi(t,x) &\approx& \Theta (-x-1/t+2/t_c-\pi/2) \Theta( x-\alpha /t -(1-\alpha)/t_c+\pi/2) \phi_{II} (t,x) \nonumber \\[-.3cm]
\\[-.3cm]
&+& \Theta (x+1/t-2/t_c+\pi/2) \Theta(-x-1/t+\pi/2) \Theta(x-1/t+\pi/2) \phi_I (t,x) \nonumber,%
\eeq%
where $\phi_{II}$ is the solution from Mathematica and $\phi_I = -{\mu \over 3} \ln (t \cos x)$.

\section{The Spherical Harmonics} \label{app-ylm}
The spherical harmonics are given by%
\beq%
Y_{lm}(\n) = \sqrt{ { (2l+1) \over 4 \pi} { (l-m)! \over (l+m)! } } P_{lm} (\cos \theta) e^{im\phi} ,%
\eeq%
where%
\beq%
P_{lm} (x) = (-1)^m (1-x^2)^{m/2} {d^m \over dx^m} P_l (x)%
\eeq%
are associated Legendre polynomials. In particular the spherical harmonics are chosen such that%
\beq%
Y_{lm} ^* &=& (-1)^m Y_{l,-m} , \nonumber \\[-.3cm]
\\[-.3cm]
\int d^2 \n \, Y_{lm} (\n) Y_{l' m'} ^* &=& \delta_{l l'} \delta_{m m'} .\nonumber%
\eeq%

\bibliographystyle{utphys}

\bibliography{bubble}

\end{document}